# Anticipation in Social Systems:
# the Incursion and Communication of Meaning

Loet Leydesdorff & Daniel M. Dubois
Amsterdam School of Communications Research (ASCoR), University of Amsterdam, Kloveniersburgwal 48, 1012 CX Amsterdam, The Netherlands.
loet@leydesdorff.net ;
http://www.leydesdorff.net

Centre for Hyperincursion and Anticipation in Ordered Systems, CHAOS asbl., Institute of Mathematics B37, University of Liège, Grande Traverse 12, B-4000 Liège, Belgium.
Daniel.Dubois@ulg.ac.be ;
http://www.ulg.ac.be/mathgen/CHAOS

**Abstract**
In social systems, meaning can be communicated in addition to underlying processes of the information exchange. Meaning processing incurs on information processing with hindsight, while information processing recursively follows the time axis. The sole assumption of social relatedness as a variable among groups of agents provides sufficient basis for deriving the logistic map as a first-order approximation of the social system. The anticipatory formulation of this equation can be derived for both anticipation in the interaction term and in the aggregation among subgroups. Using this formula in a cellular automaton, an observer is generated as a reflection of the system under observation. The social system of interactions among observations can improve on the representations entertained by each of the observing systems.
**Keywords:** social systems, anticipation, observer, meaning, logistic map

## 1 Introduction

Rosen (1985) defined an anticipatory system as a system that contains a model of the system itself. For example, a biological system can use this internal representation for anticipatory adaptation, that is, to predict the survival value of the system among its possible manifestations at a next moment in time. Dubois (2000) distinguished between weak anticipation, that is, when systems use a model of themselves for computing future states, and strong anticipation, that is, when the system uses itself for the construction of its future states. In the latter case anticipation is no longer similar to prediction.

In this paper we argue that the social system can be considered as anticipatory in the strong sense: this system constructs its future by providing the expected information content of the distribution of events with meaning. The anticipations can be communicated among the agents in a next-order network that feeds back on the information-processing network. However, meaning is provided with hindsight (that is, *a posteriori*), and therefore meaning processing also feeds back on the time axis *within* the system (Luhmann, 1984; Leydesdorff, 2001a).

The meaning processing thus adds a reflexive layer of communication to the information processing in social systems. (The interaction between these two layers then produces meaningful information.) While the historical configurations of social systems change in a forward mode in terms of both uncertainty and meaningful information, the information processing is internally subject to reflections *ex post*. The meaning processing reduces the uncertainty contained in the distributions of first-order networks locally.

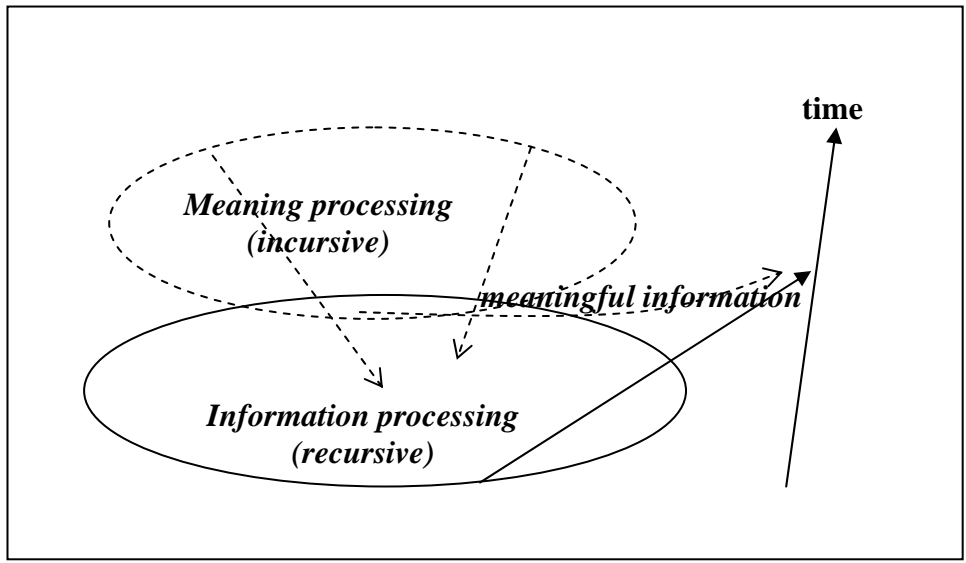

**Figure 1:** The incursive processing of meaning interacts with the recursive processing of information and the result is the localized production of meaningful information

Providing information with meaning can be considered as a selective operation. Some uncertainty in the information processing is discarded as noise, and other uncertainty is identified as 'meaningful information.' Thus, the meaning processing structures the information processing. Meaning processing continuously reflects on the system of information processing under observation. The two processes can be considered 'structurally coupled' (Maturana, 1978): at the level of the social system the one process cannot operate without the other. Biological systems can provide meaning to information, but cannot exchange the meanings thus generated among themselves.[1]

The reflections provide us with mirror images, but from potentially different perspectives. When the reflections can again be communicated, they are recursively built into the historical (that is, forward) development of the social system. The exchange of meaning adds globally to the information processing by distinguishing the meaningful information from the noise in terms relevant for the reproduction of the social system (Urry, 2003). Each communication leads to new communications, and thus the social system continuously reconstructs the order of expectations from a hindsight perspective by operating on the layers that it has generated historically.

---

[1] The psychological system is expected not only to process meaning, but also to generate identity. Unlike the social system, the dynamics at this level can under certain conditions become historically fixed.

The purpose of this paper is: (1) to model the social as a system containing strong anticipation, (2) to prove the equations, and (3) to show how this system works by using simulations. In the next section, we first derive the anticipatory formulation of the logistic equation for aggregation and interaction among subgroups of the social system. Using these algorithms, simulations enable us subsequently to generate a reflexive observer *within* an information-processing system. Cellular automata will be used for the visualization of how social systems operate (Leydesdorff, 2001b and 2002).

## 2   The Specification of a Social System

The double-layeredness of the operation of a social system processing both information historically and meaning with hindsight can be described by using the incursive formulation of the logistic map as proposed by Dubois (1998):

$$x(t) = a\, x(t-1)\, \{1 - x(t)\} \tag{1a}$$
$$\text{or:} \quad x(t+1) = a\, x(t)\, \{1 - x(t+1)\} \tag{1b}$$

For example, the price of a commodity can be considered as its expected value on the market. The price codifies the value of a commodity in economic terms. The anticipatory formulation of the logistic curve appreciates that the price has both an intrinsic value and is reflected in a feedback of the market system. The intrinsic factor stems from the historical production process, while the feedback from the market originates in the present on the basis of the dynamics of current supply and demand.

The use of the traditional—that is, only forward—format of this equation is ill-advised, since the two subdynamics of production and diffusion are then not sufficiently distinguished in terms of the dynamics over time. Production proceeds historically along the time axis—for example, building on previous generations of a technology—while diffusion takes place under competitive conditions in the present. The selection mechanism (that is, the market) can thus be considered as an evolutionary feedback on the historical development (Andersen, 1994; Leydesdorff & Van den Besselaar, 1998; Nelson & Winter, 1982).

The techno-economic system can be modeled using this anticipatory version of the logistic equation. The recursion on x(t-1) in the left-hand term of eq. 1a represents the axis of historical development of the technology. The system additionally selects in the present upon the development as declared in the right-hand term of the equation. The selection pressure prevailing in the present is analytically independent of the previous state of the system that produced the variation. Thus, the two mechanisms interact as subdynamics of the social system.

## 2.1 The definition of a social system

Let us first consider two groups y and x in a social system. The behavior of these groups can be described by the following equations:

$$dy(t)/dt = -ax(t)y(t) + bx(t) \tag{2a}$$
$$dx(t)/dt = +ax(t)y(t) - bx(t) \tag{2b}$$

Let us furthermore assume that $x(t) + y(t) =$ Constant $= C$; for example, $C = 1$. The parameter $b$ can also be taken as $b = 1$, without losing any generality.

The sociological interpretation of this system of equations is as follows: the two groups $x(t)$ and $y(t)$ interact with an interaction given by the product of the two populations $x(t)y(t)$ at the rate $a$. The interaction between the two groups x and y can be considered as the sociability factor of the population.

For example, the group y may represent isolated persons, and the group x persons entertaining relations with each other. When there is interaction, eq. 2a specifies that the number of isolates decreases with the interaction term $(-ax(t)y(t))$, while the chance of interaction increases with the number of persons already pertaining relations $(+bx(t))$. In eq. 2b, similarly, the number of already related persons increases because of this interaction $(+ ax(t)y(t))$, but the chance of further relatedness decreases $(-bx(t))$ when a larger part of the population has already been related.[2]

For $a = 0$, $x(t)$ becomes zero and the whole population would constituted of isolated persons;[3] $y(t)$ is in this case equal to 1 because $x + y = 1$. The parameter $a$ can thus also be considered as a measure of the sociability of the population (cf. Dubois & Sabatier, 1998).

## 2.2 Derivation of the logistic equation

Assuming $b = 1$ in eqs. 2a and 2b, the corresponding discrete system can be formulated as follows:

$$y(t+1) = y(t) - ax(t)y(t) + x(t) \tag{3a}$$
$$x(t+1) = x(t) + ax(t)y(t) - x(t) = ax(t)y(t) \tag{3b}$$

Because
$$y(t+1) + x(t+1) = y(t) + x(t) = \text{Constant} = 1 \tag{4a}$$
$$y(t) = 1 - x(t) \tag{4b}$$

The logistic map is obtained by replacing eq. 4b into eq. 3b:

---

[2] The model is known in the literature as the SIS model of Bailey (1957) which simplifies on the classical SIRS model for the spread of infectious diseases by Kermack & McKendrick (1927).

[3] The other root of eq. 2b is that x = -2. Since x + y = 1, y = 3 in this case. This solution does not have an obvious interpretation.

x(t+1) = *a*x(t){1 − x(t)}

As is well known, this model generates chaotic behavior for *a* = 4.

**2.3   The anticipatory version of the model**

One can consider two anticipatory versions of the above model in the case of a social system. First, one may expect the grouping process itself to contain anticipation. For example, isolated individuals may consider whether it is to their advantage to enter into relations. Second, one can assume that the interaction term between the two groups x and y contains anticipation. We will now first prove that both assumptions lead to the anticipatory version of the logistic map as specified above in eq. 1b.

2.3.1   Anticipation in the development of y

In general, the anticipatory model is an analytical result of the backward evaluation of the differential equation in discrete time:

$$x(t - \Delta t) = x(t) - \Delta t \, f(x(t)) \tag{5}$$

Applied to eq. 3a this leads to the following model:

$$y(t+1) = y(t) - ax(t+1)y(t+1) + x(t+1) \tag{6a}$$

In this model, y—that is, the grouping of isolated persons—contains anticipation since the term is operating upon both its previous and its present state at the same time. Without the further assumption of anticipation in the interaction (see the next section), the development of x(t) remains unchanged (as in eq. 3b):

$$x(t+1) = x(t) + ax(t)y(t) - x(t) = ax(t)y(t) \tag{6b}$$

Because of the anticipation in y, however, y in this case relates to the next state of x, and therefore:

$$\begin{aligned} y(t) + x(t+1) &= \text{Constant} = 1 \\ y(t) &= 1 - x(t+1) \end{aligned} \tag{7}$$

By putting eq. 7 into 6b, one obtains:

$$x(t+1) = ax(t)(1 - x(t+1)) \tag{8}$$
*Q.e.d.*

### 2.3.2 Incursion in the interaction between x with y

Let us now assume that the interaction term between x and y contains the source of anticipation:

$$y(t+1) = y(t) - ax(t)y(t+1) + x(t) \tag{9a}$$
$$x(t+1) = x(t) + ax(t)y(t+1) - x(t) = ax(t)y(t+1) \tag{9b}$$

For analytical reasons, one can also write the interaction term as a difference equation in relation to its previous state, as follows:

$$x(t)y(t+1) = x(t)y(t) + x(t)\{y(t+1) - y(t)\} \tag{10a}$$
$$y(t+1) = y(t) + \{y(t+1) - y(t)\} \tag{10b}$$

In other words, the anticipatory interaction depends on a supplementary factor given by the derivative of y. Since both terms are thus implied in the anticipation:

$$y(t+1) + x(t+1) = \text{Constant} = 1$$

and therefore:

$$y(t+1) = 1 - x(t+1) \tag{11}$$

By replacing eq. 11 into eq. 9b, one obtains again:

$$x(t+1) = ax(t)\{1 - x(t+1)\} \tag{12}$$
*Q.e.d.*

In summary, the introduction of anticipation into a very basic model of the social system can be shown to lead to the anticipatory formulation of the logistic equation. First, we argued for using this equation on theoretical grounds, and in this section we have derived this model of the social system from assumptions about the possible contingencies between two subpopulations (Parsons, 1968).

## 3 The Simulations

Social systems are based on exchange relations. In other words, social systems are distributed by their very nature. Cellular automata enable us to display the dynamics of multi-agent systems in terms of colours on the screen. Each point (x, y) on the screen can be considered as an agent which relates—or not—to other agents. Different colours can be used to indicate the phenotypical state of the various agents over time. In addition to this visualization, the value of each pixel can be mapped for computational purposes in an array (x, y) with the size of the screen.

In order to enhance the transparency, we formulate the simulation models in standard BASIC.[4] For example, the array is defined in line 40 of Table 1 so that it can contain a representation of the screen in CGA-mode (200 x 320 pixels). The CGA-mode (line 10) was chosen in order to take full advantage of the visibility of the effects on the screen. A pixel—representing an agent—is selected randomly in lines 110 and 120.

**Table 1:** Incursion and recursion in lines 140 and 150, respectively.

```
1    CLS : LOCATE 10, 10: INPUT 'Parameter value'; a
2    IF a > 4 THEN a = 4                        ' prevention of overflow

10   SCREEN 7: WINDOW (0, 0)-(320, 200): CLS
20   RANDOMIZE TIMER
30   ' $DYNAMIC
40   DIM scrn(321, 201) AS SINGLE
50   FOR x = 0 TO 320
60     FOR y = 0 TO 200
70       scrn(x, y) = .1: PSET (x, y), (10 * scrn(x, y))
80     NEXT y
90   NEXT x

100 DO
110   x = INT(RND * 320)
120   y = INT(RND * 200)
130   IF y > 100 GOTO 140 ELSE GOTO 150     ' split of screens
140      scrn(x, y) = a * scrn(x, y) / (1 + a * scrn(x, y)): GOTO 160
150      scrn(x, y) = a * scrn(x, y) * (1 - scrn(x, y))
160   PSET (x, y), (10 * scrn(x, y))
170   LOOP WHILE INKEY$ = ''
180 END
```

On the assumption that a social system contains the two layers of information and meaning processing, we use the logistic equation in the forward mode for the historical information processing, and in the anticipatory formulation for providing meaning to the information processing. In the program (Table 1) and the corresponding Figure 2, the screen is accordingly split into two halves (line 130).[5]

In the lower half, the results of the logistic evaluation of the corresponding array value (line 150) are brought to the screen in line 160. In the upper half, the anticipatory version of the logistic equation is used for the evaluation, and the result of this evaluation is also depicted in line 160. The analytical rewrite of the logistic equation in the format used in line 140 is provided in Table 2 (Dubois, 1998). The code in the first line enables the user to choose the parameter value for *a* interactively.[6]

**Table 2:** Analytical rewrite of the anticipatory formulation of the logistic equation

$$x(t) = ax(t-1)(1 - x(t)) \qquad (1a)$$
$$x(t) = ax(t-1) - ax(t-1)\,x(t)$$
$$x(t) + ax(t-1)\,x(t) = ax(t-1)$$
$$x(t)(1 + ax(t-1)) = ax(t-1)$$
$$x(t) = ax(t-1) / (1 + ax(t-1))$$

---

[4] The programs can be adapted for higher or commercial versions of Basic, and for other languages. In Visual Basic the programs formulated in this paper can be imported as subroutines.
[5] An interactive version of the simulations can be retrieved at http://www.leydesdorff.net/casys03 .
[6] In order to prevent overflow while running this model, values of the parameter *a* larger than 4 are reset to *a* = 4 (in line 2).

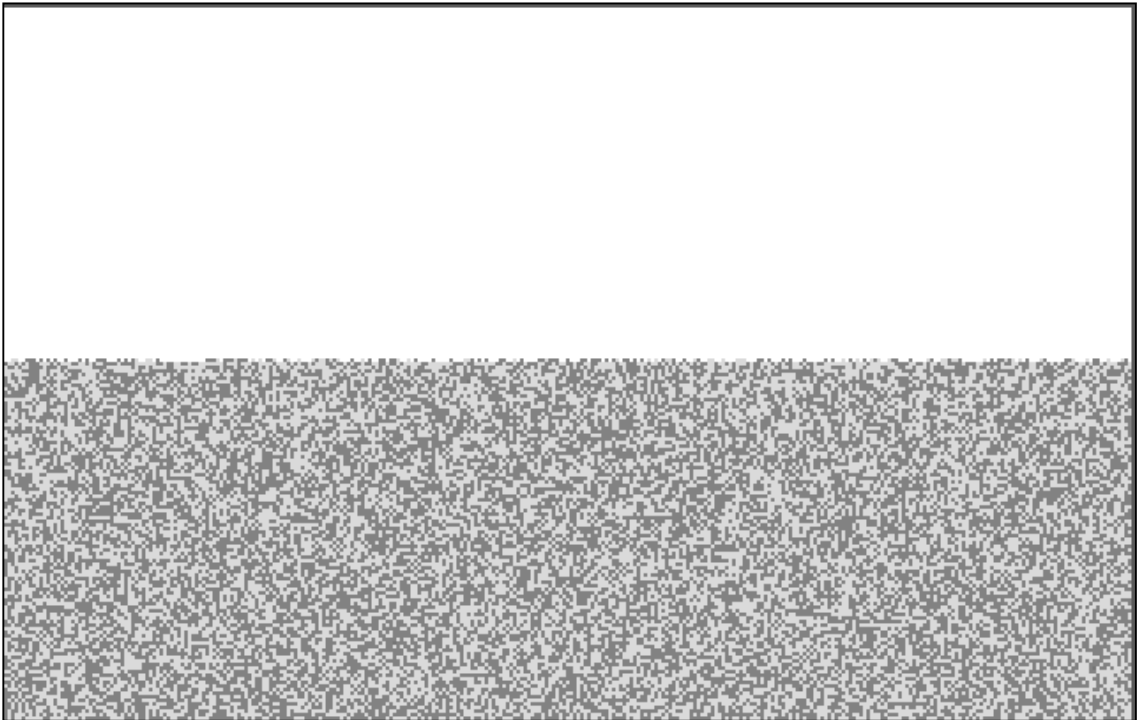
**Figure 2:** Upper half of the screen incursive, lower half recursive; $a$ = 3.1

Figure 2 shows that the incursive simulation leads to a transition, while the representation of the recursive system in the lower half of the screen exhibits the bifurcation as expected for the value of $a$ = 3.1. The incursive model converges to a stable state (in this case, exhibited as white) because the $\mathrm{Lim}_{x\to\infty} \{ax/(1+ax)\} = 1$.

In the next simulation (Table 3) we combine the two subdynamics of incursion and recursion into one single screen. In this model the choice of the incursive or recursive routines is randomly assigned in line 130, but the screen and the array values are no longer split according to the value of the vertical coordinate.

**Table 3:** Incursion and recursion alternating randomly, but using the same data set

```
[...]
100 DO
110    x = INT(RND * 320)
120    y = INT(RND * 200)
130    IF RND > .5 GOTO 140 ELSE GOTO 150
140       scrn(x, y) = a * scrn(x, y) / (1 + a* scrn(x, y)) : GOTO 160
150       scrn(x, y) = a * scrn(x, y) * (1 - scrn(x, y))
160    PSET (x, y), (10 * scrn(x, y))
170 LOOP WHILE INKEY$ = ''
180 END
```

When the incursive model operates within a recursive system of which it is also a part, the incursive routine tends to reduce the uncertainty produced by the recursive one, since incursion drives towards a transition in the long run because of the noted limit. The transition is visible on screen as a trend toward a dominant colour, but this transition is not achieved because the incursive routine is continuously disturbed by the recursive one. The system therefore remains in transition.

Note that from an (historical) actor perspective the incursive transition operates as a latent attractor. The longer-term prevalence of incursion over recursion, however, demonstrates the importance of accounting for expectations in models of the evolution of social systems when both subdynamics can be expected to play a role in the system. The emerging layers of social coordination, that is, the communication of meaning, can be expected to provide additional stabilities because of the selective capacity of the implied coding.

For example, instantaneous selections can be selected recursively for historical stabilization over time. This occurs in processes like institutionalization. By using incursion and therefore time as another degree of freedom, some historical stabilization can be selected for meta-stabilization or globalization. This next-order level remains pending as selection pressure on the historical manifestation.

## 4 The Generation of an Observer

Can the result of the interacting dynamics of a complex system that contains both incursion and recursion also be decomposed into an observing and an observed subsystem? In the model exhibited in Table 4 and Figure 3, the two routines of 'observed' and 'observing' are decomposed so that an observer is generated by using the incursive routine.[7] The upper half of the screen is reserved for exhibiting the results of the incursive observations of the lower half of the screen, while the lower half is based on the recursive routines and therefore exhibits the historical development of the observed system.

In order to generate an observable structure at each moment in time, a network effect was added to the observed system (in lines 110-120 of Table 4). This network effect spreads the update in the lower-level screen in the local (Von Neumann) neighbourhood of the affected cell. (The Von Neumann environment is defined as the cells above, below, to the right, and to the left of the effect.) The network effect enables us to appreciate on the screen the development of both the observed system and the relative quality of the observation depicted in the upper half of the screen. Note that the network effect structures the system at each moment in time and locally, whereas incursion and recursion are defined over the time axis of the system, that is, as an operation at the system's level.

Henceforth, we use the full range of 16 colours available in the BASIC palette in order to provide more details on the screen. This is achieved by changing the decimal base of the above simulations to the basis of 16 (in line 43 of Table 4). Whenever necessary normalizations of the formulas for incursion and recursion are added by dividing again by 16 (for example, in lines 160 and 180).

---

[7] The subsystem entertaining the model of the system in the present state can be considered as an endogenous observer of the system's history. Endogenous means here that this observer remains a result of the network in which the observer effect is generated (Maturana, 1978). One can consider this observer as an incursive subroutine of the complex system. Note that the metaphor is still biological because this observer is not positioned refexively in a (next-order) communication among observers (Leydesdorff, 2000; Maturana & Varela, 1980). The observer remains completely embedded and follows the development in the observed system.

**Table 4:** The generation of an observer by using incursion

```
1    CLS: LOCATE 10, 10: 'Parameter value for the recursion (a)'; a
2         LOCATE 11, 10: 'Parameter value for the incursion (b)'; b
3    IF a > 4 THEN a = 4

10   SCREEN 7: WINDOW (0, 0)-(320, 200): CLS
11   LINE (1, 100)-(320, 100)
20   RANDOMIZE TIMER

30   ' $DYNAMIC
40   DIM scrn(321, 201) AS INTEGER
50   FOR x = 0 TO 320
60      FOR y = 0 TO 200
70         scrn(x, y) = INT(RND * 16)       ' change to 16 colours
80         PSET (x, y), scrn(x, y)          ' (see note 4)
90      NEXT y
100  NEXT x

110  DO
120     y = INT(RND * 200)
130     x = INT(RND * 320)
140     IF (x = 0 OR y = 0) GOTO 220        ' prevention of network errors
150     IF y > 100 GOTO 160 ELSE GOTO 180
160        scrn(x, y) = b * scrn(x,y-100) / (1 + b * (scrn(x,y-100) / 16))
170        GOTO 210                         ' paint upper screen
180        scrn(x, y) = a * (scrn(x, y) * (1 - (scrn(x, y)) / 16))
              ' spread new value in the Von Neumann environment
190        scrn(x + 1, y) = scrn(x, y): scrn(x - 1, y) = scrn(x, y)
200        scrn(x, y + 1) = scrn(x, y): scrn(x, y - 1) = scrn(x, y)
210     PSET (x, y), ABS(scrn(x, y))
220  LOOP WHILE INKEY$ = ''
230  END
```

Whereas the incursive and the recursive routines operated on the same initial configurations as in the model provided in Table 3, the feedback relation between the two systems changes continuously in this model. In this model, the two parameters for recursion (*a*) and incursion (*b*) can also be varied independently. A random attribution is decisive (in line 150) for whether the recursive or the incursive routine is entered. However, the incursive routine (line 160) operates on the value of the corresponding array element in the lower half of the screen by evaluating scrn(x, y-100). The result of this evaluation is attributed to the upper half of the screen and to the corresponding array value of scrn(x, y). The effect is that an observer is generated as exhibited in Figure 3 (*a* = 3.2 and *b* = 3.2).

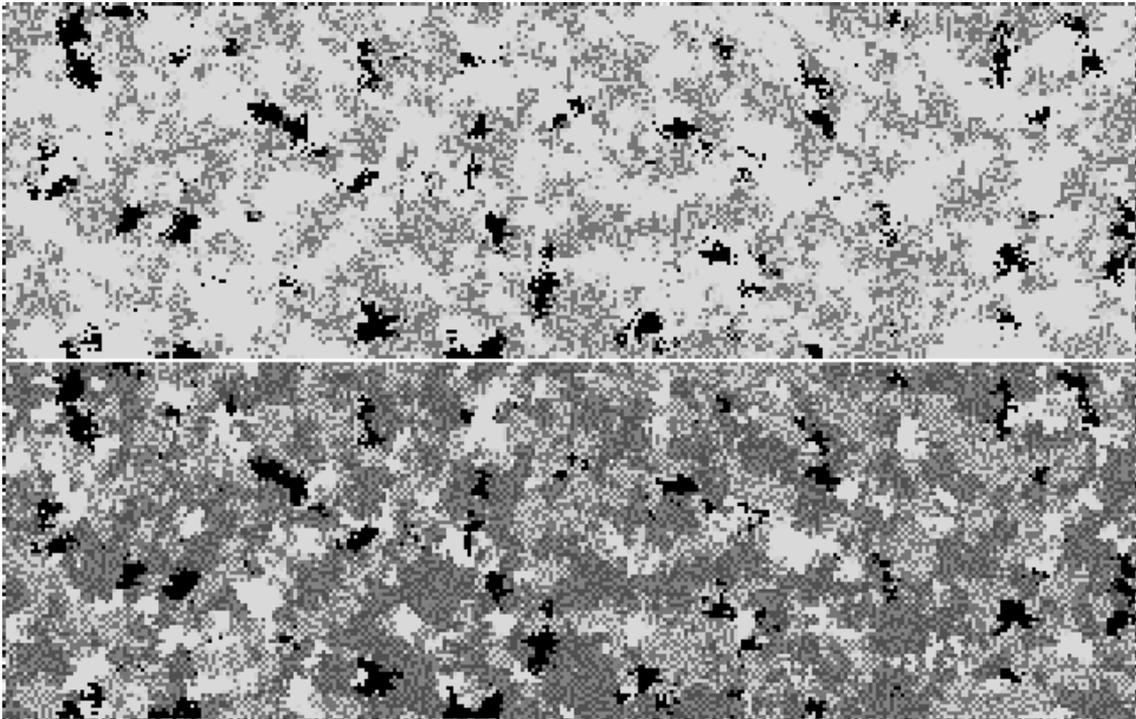
**Figure 3:** Incursion and recursion with different parameter values produce observers with potentially different positions and corresponding blind spots (*a* = 3.2 and *b* = 3.2).

By changing the parameter of the incursion, one can change the window of observation of an observer. High values for the incursion parameter (*b*) drive the observing system into a more homogeneous state (because of the above noted limit transition in the formula), while higher values of the recursive parameter (*a*) drive the historically developing system towards more chaotic bifurcations.

## 5   Observing the Observers

The possibility of generating observers with the different qualities of their respective observations raises the question of the possibility of interaction among the observers, for example, when the observers observe each other's observations. Human observers can additionally interact by using more sophisticated mechanisms like a human language or symbolic media for the communication (Luhmann, 1982, 1997; Parsons, 1963a; 1963b). This further extension is the subject of a next study, but some expectations can be anticipated by showing the results of a single simulation.

Figure 4 exhibits the results of two observers with different parameters *b* and *c* observing the recursive development in the lower left screen (with parameter *a*). The two observers additionally observe each other's observations, and the lower right quadrant is used for the exhibition of the results of interactive and aggregative combinations of these observations.

The lower right quadrant shows a representation of the observed system in the lower left quadrant that is richer in detail than either of the observations by the individual observers in the two upper-half quadrants. It should be remembered that the two incursive observers operate at random frequencies with different parameters. Consequently, an interaction among the observations contains a dynamic uncertainty that may represent elements of the originally observed system which are lost in the individual reflections, while the latter focus on the observable structure and thus reduce the complexity. The aggregation or averaging of the different observations can be expected to lead to uncertainty in the delineations at each moment in time; the interaction of the reflections opens a phase space of possible reconstructions of the observed system.

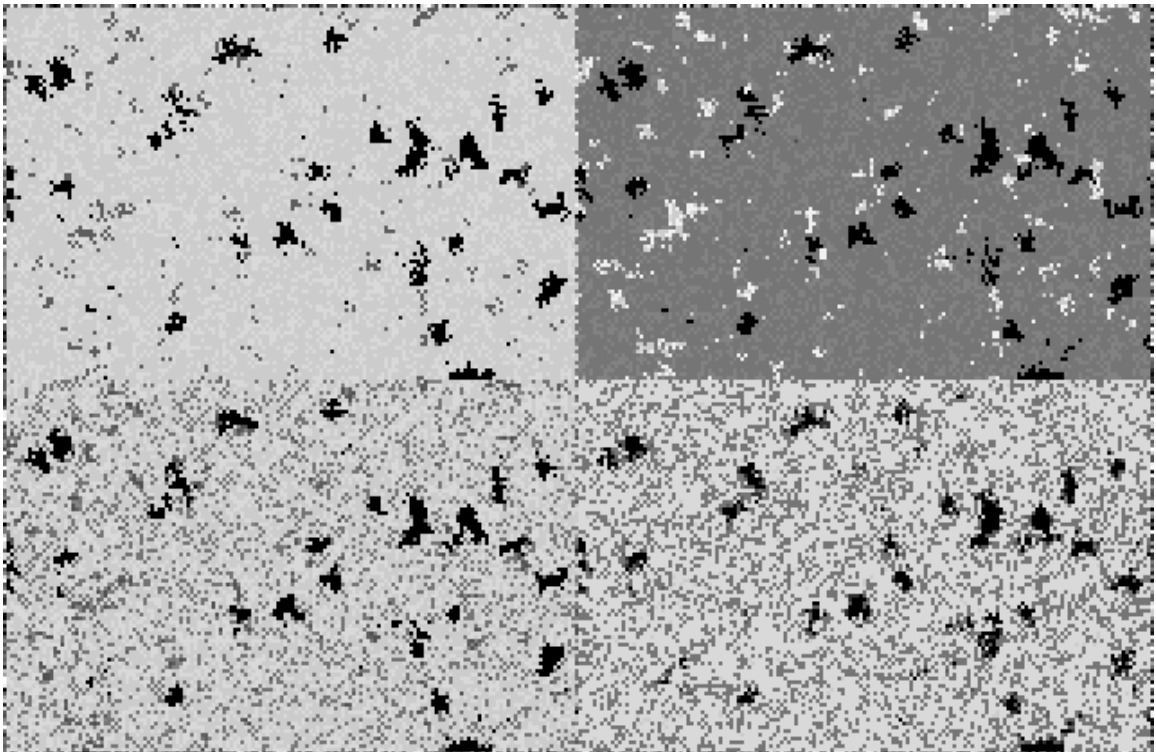

**Figure 4:** A non-linear combination of observations by two observers in the lower right quadrant

## 6 Conclusions

We have argued that the social system can be considered as an anticipatory system in the strong sense of constructing its own future. It does so by reconstructing its past in the present. Because this reconstruction is functional to the progressive development of this system, the system can be expected to differentiate increasingly and a manifold of meanings can be entertained. The ones which are again selected, are circulated as information in a system that thus remains under continuous reconstruction.

The exchange processes of meaning constitute a layer on top of the historical exchanges of information. This double layered process was modeled using the traditional (recursive) formulation of the logistic map for the historical process and the anticipatory formulation for the evolutionary process that changes the historical process in a distributed mode.

The appropriateness of the approach was derived from the sole assumption of sociability in the social process. However, the agents should additionally be competent to communicate in terms of exchanging both meaning and uncertainty. Human language can perhaps be considered as the evolutionary achievement that enables us to entertain these communicative competences in relation to each other, but without the need for a harmonic resolution.

## References

Andersen, E. S. (1994). Evolutionary Economics: Post-Schumpeterian Contributions. London: Pinter.

Bailey, N. T. J. (1957). The Mathematical Theory of Infectious Diseases and Its Applications. London: Charles Giffin & Co.

Dubois, D. M. (1998). Computing Anticipatory Systems with Incursion and Hyperincursion. Computing Anticipatory Systems: CASYS'97 – First International Conference. Edited by Daniel M. Dubois, Published by The American Institute of Physics, AIP Conference Proceedings 437, pp. 3-29.

Dubois Daniel M. (2000) Review of Incursive, Hyperincursive and Anticipatory Systems - Foundation of Anticipation in Electromagnetism. Computing Anticipatory Systems: CASYS'99 - Third International Conference. Edited by Daniel M. Dubois, Published by The American Institute of Physics, AIP Conference Proceedings 517, pp. 3-30.

Dubois, D. M., and Ph. Sabatier (1998). Morphogenesis by Diffuse Chaos in Epidemiological Systems. Computing Anticipatory Systems: CASYS'97 – First International Conference. Edited by Daniel M. Dubois, Published by The American Institute of Physics, AIP Conference Proceedings 437, pp. 295-308.

Kermak, W. O., & A. G. McKendrick. (1927). Contribution to the Mathematical Theory of Epidemics. Proc. Roy. Soc. A, 115, 700-721.

Leydesdorff, L. (2000). Luhmann, Habermas, and the Theory of Communication. Systems Research and Behavioral Science, 17, 273-288.

Leydesdorff, L. (2001a). A Sociological Theory of Communication: The Self-Organization of the Knowledge-Based Society. Parkland, FL: Universal Publishers; at <http://www.upublish.com/books/leydesdorff.htm >.

Leydesdorff, L. (2001b). Technology and Culture: The Dissemination and the Potential 'Lock-in' of New Technologies. Journal of Artificial Societies and Social Simulation, 4(3), Paper 5, at <http://jasss.soc.surrey.ac.uk/4/3/5.html>.

Leydesdorff, L. (2002). The Complex Dynamics of Technological Innovation: A Comparison of Models Using Cellular Automata. Systems Research and Behavioral Science, 19(6), 563-575.

Leydesdorff, L., & P. v. d. Besselaar. (1998). Technological Development and Factor Substitution in a Non-Linear Model. Journal of Social and Evolutionary Systems, 21, 173-192.

Luhmann, N. (1982). Liebe als Passion. Frankfurt a.M.: Suhrkamp.

Luhmann, N. (1984). Soziale Systeme. Grundriß einer allgemeinen Theorie. Frankfurt a. M.: Suhrkamp.

Luhmann, N. (1997). Die Gesellschaft der Gesellschaft. Frankfurt a.M.: Suhrkamp.

Maturana, H. R. (1978). Biology of language: the epistemology of reality. In Psychology and Biology of Language and Thought. Essays in Honor of Eric Lenneberg. Edited by G. A. Miller & E. Lenneberg. New York: Academic Press, pp. 27-63.

Maturana, H. R. & F. Varela (1980). Autopoiesis and Cognition: The Realization of the Living. Boston: Reidel.

Nelson, R. R. & S. G. Winter (1982). An Evolutionary Theory of Economic Change. Cambridge, MA: Belknap Press of Harvard University Press.

Parsons, T.S. (1968). Interaction: I. Social Interaction. The International Encyclopedia of the Social Sciences. Edited by D. L. Sills. New York: McGraw-Hill, Vol. 7, pp. 429-441.

Parsons, T. S. (1963a). On the Concept of Political Power,. Proceedings of the American Philosophical Society 107(3), 232-262.

Parsons, T. S. (1963b). On the Concept of Influence. Public Opinion Quarterly 27 (Spring), 37-62.

Rosen, R. (1985). Anticipatory Systems. Oxford, etc.: Pergamon Press.

Urry, J. (2003). Global Complexity. Cambridge, UK: Polity.